\documentclass[prd,twocolumn,superscriptaddress,showpacs,floatfix,%
preprintnumbers, nofootinbib,hyperref]{revtex4}              
\usepackage{graphicx}
\usepackage{rotating}
\usepackage{epsfig}
\usepackage[usenames,dvipsnames]{xcolor}

\usepackage{color}

\def\he4{$^4$He}
\def\h2{$^2$H}

\begin{document}

\preprint{IPPP/14/ 87; DCPT/14/ 174}

\title{Collisional production of sterile neutrinos via secret interactions\\
and cosmological implications}
\author{Alessandro Mirizzi} 
\affiliation{II Institut f\"ur Theoretische Physik, Universit\"at Hamburg, Luruper Chaussee 149, 22761 Hamburg, Germany}
\author{Gianpiero Mangano}
\affiliation{Istituto Nazionale di Fisica Nucleare - Sezione di Napoli, Complesso Universitario di Monte S. Angelo, I-80126 Napoli, Italy}
\author{Ofelia Pisanti}
\affiliation{Istituto Nazionale di Fisica Nucleare - Sezione di Napoli, Complesso Universitario di Monte S. Angelo, I-80126 Napoli, Italy}
 \affiliation{Dipartimento di Fisica, Universit{\`a} di Napoli Federico II, Complesso Universitario di Monte S. Angelo, I-80126 Napoli, Italy}
 \author{Ninetta Saviano} 
\affiliation{Institute for Particle Physics Phenomenology, Department of Physics,
Durham University,\\ Durham DH1 3LE, United Kingdom}

\date{\today}

\begin{abstract}
Secret interactions   among sterile neutrinos  have been recently proposed as an escape-route to reconcile eV sterile neutrino hints from short-baseline
anomalies with cosmological observations. In particular models with coupling
$g_X \gtrsim 10^{-2}$ and gauge boson mediators $X$ with $M_X \lesssim 10$~MeV lead to large matter potential suppressing the sterile neutrino production 
 before the neutrino decoupling.
 With this choice of parameter ranges, big bang nucleosynthesis is left unchanged and gives no bound on the model. However, we show that at lower temperatures when active-sterile oscillations are no longer
matter suppressed,  sterile neutrinos are still in a collisional regime, due to their
secret self-interactions. The  interplay between  vacuum oscillations and
 collisions  leads to a  scattering-induced decoherent production of sterile neutrinos with a fast rate.
This process is  responsible for
a flavor equilibration among the different neutrino species. 
We explore the effect of this large sterile neutrino population on cosmological observables.
We find that a signature of strong secret interactions would be a reduction of the effective number 
of neutrinos  $N_{\rm eff}$ at matter radiation equality down to 2.7. Moreover, for $M_X \gtrsim g_X$~MeV  sterile neutrinos would be free-streaming before becoming non-relativistic and they
would affect the large-scale structure power spectrum.  As a consequence, for this range of parameters we   find a tension of a eV mass sterile state with cosmological neutrino mass bounds.

\end{abstract}

\pacs{14.60.St, 
	   14.60.Pq, 
		98.80.-k 
}  

\maketitle

\section{Introduction}
In the recent years there has been a renewed attention on
eV sterile neutrinos, suggested by anomalies found in the short-baseline neutrino experiments (see~\cite{Abazajian:2012ys}
for a recent review).
In this  context, it has been realized that these light sterile states would be efficiently 
produced in the early universe by oscillations with active neutrinos, leading to a conflict with different
cosmological observations~\cite{Hannestad:2012ky,Hamann:2011ge}. 
This problem motivated the investigation of different mechanisms to suppress the sterile neutrino
 thermalization (see, e.g., ~\cite{Hannestad:2012ky,Mirizzi:2012we,Saviano:2013ktj,Ho:2012br}).
In~\cite{Hannestad:2013ana,Dasgupta:2013zpn} it has been recently proposed a novel suppression mechanism
based on the introduction of secret interactions  
among sterile neutrinos, mediated by a massive gauge boson $X$, with $M_X \ll M_W$ 
(see~\cite{Archidiacono:2014nda}
for the case of sterile neutrinos interacting with a ligh pseudoscalar).
Indeed, these secret interactions  would generate a large matter term in the sterile neutrino sector, which lowers the effective
neutrino in-medium mixing angle. 

However, as the matter potential declines, a  resonance is eventually encountered. Sterile neutrinos would 
be produced by a combination of (damped)  Mikheyev-Smirnov-Wolfenstein
 (MSW)-like  resonant flavor conversions  among active and sterile
neutrinos~\cite{Matt},  and the non-resonant processes associated with the secret collisional effects~\cite{Kainulainen:1990ds}.
It particular,  assuming for secret interactions a coupling costant $g_X \gtrsim 10^{-2}$ and masses of the mediator
$M_X \gtrsim {\mathcal O}(10)$~MeV, the sterile neutrino production would occur  at $T \gtrsim 0.1$~MeV, with non trivial
consequences on big bang nucleosynthesis (BBN)~\cite{Hannestad:2013ana}. In this context, in~\cite{Saviano:2014esa} it has been shown that BBN observations would lead
to severe constraints on the parameter space of the model, 
reducing the range which satisfies cosmological bounds. 

In~\cite{Dasgupta:2013zpn} much lighter bosons were considered.
 The advantage of this choice seems twofold. On one hand 
the matter potential would be so strong to inhibit any sterile neutrino production
during BBN, thus evading the corresponding constraints.
 Moreover,  
 if the new interaction mediator $X$  couples not only to sterile neutrinos but also to dark matter particles, 
 for such small masses it might also possibly relieve some of the small-scale structure problems of the cold dark matter scenario~\cite{Dasgupta:2013zpn,Bringmann:2013vra,Ko:2014bka}
(nonstandard interactions were also introduced 
 to alleviate these problems in~\cite{Boehm:2000gq,Boehm:2003hm,Serra:2009uu,Aarssen:2012fx,Boehm:2014vja,Chu:2014lja}
and in the references therein). 
Secret interactions among sterile neutrinos, mediated by very light (or even massless) pseudoscalars,
and their connection with dark matter was explored in~\cite{Archidiacono:2014nda}.
 
The aim of this paper is to show that also for these small masses a large sterile neutrino production 
is  unavoidable at $T \ll 0.1$~MeV, 
when the matter potential becomes smaller than the vacuum oscillation term. 
Indeed, the small vacuum oscillations  act as  \emph{seed} for 
a \emph{ scattering-induced decoherent production} of sterile neutrinos,
associated with
the secret self-interactions in the sterile sector. This process is very rapid and
  leads to a quick flavor equilibration among the active and 
the sterile neutrino species
 after the active neutrino decoupling. 
We will explore the consequences of this large  sterile neutrino abundance on 
cosmological observables, namely the effective number of neutrinos $N_{\rm eff}$ at matter radiation equality and recombination and
the sterile neutrino mass bounds.

The paper is organized as follows.
 In Section II we present an overview of the flavor evolution for the active-sterile
neutrino system in the presence of secret interactions.
In Section III we discuss the  scattering-induced decoherent production of sterile neutrinos,
 associated with the
secret interactions in the post-decoupling epoch. In Section IV
we discuss the observable signatures of this scenario for very fast sterile-sterile scattering processes. We find that  active-sterile neutrino flavor conversions leads to a reduction of $N_{\rm eff}$ down to 2.7. Furthermore, for $M_X \gtrsim g_X$~MeV  sterile neutrinos would be free-streaming before they become a non relativistic species. Thus, their large number density would affect the large-scale
structure power spectrum.  In this case, we compare the sterile neutrino abundance
with the most recent cosmological mass bounds finding a tension for an
eV mass range  sterile state. 
Finally, in Section V we summarize our results and conclude.

 \section{Flavor evolution}
 We consider  a 3+1 neutrino mixing scenario, involving the three active families and 
a sterile species. 
As usual, we 
describe  the neutrino (antineutrino) system in terms of  $4\times 4$ density matrices $\rho= \rho(p)$, for the different neutrino momenta $p$.
The evolution equation  of the  density matrix $\rho$ is ruled by the kinetic equations~\cite{Sigl:1992fn,McKellar:1992ja,Mirizzi:2012we}
\begin{equation}
{\rm i}\,\frac{d\rho}{dt} =[{\sf\Omega},\rho]+ C[\rho]\, .
\label{drhodt}
\end{equation}
Assuming no neutrino asymmetry, the dynamics of antineutrinos is identical to the one of neutrinos.
The evolution in terms of the comoving observer proper time $t$ can be easily recast in function of the 
neutrino temperature $T_\nu$ 
(see~\cite{Mirizzi:2012we} for a detailed treatment). 
The first term on the right-hand side of Eq.\ (\ref{drhodt}) describes the flavor oscillations Hamiltonian, given by
\begin{eqnarray}
{\sf\Omega}&=&\frac{{\sf M}^2}{2 p} +
\sqrt{2}\,G_{\rm F}\left[-\frac{8  p}{3 }\, \bigg(\frac{{\sf E_\ell}}{M_{\rm W}^2} + \frac{{\sf E_\nu}}{M_{\rm Z}^2}\bigg)
\right] \nonumber \\
&+& \sqrt{2}\,G_{\rm X}\left[-\frac{8  p  {\sf E_s}}{3 M_{\rm X}^2}
\right]  \,\ ,
\label{eq:omega}
\end{eqnarray}
where ${\sf M}^2$ = ${\mathcal U}^{\dagger} {\mathcal M}^2 {\mathcal U}$ is the neutrino mass matrix,  written in
terms of the solar $\Delta m^2_{\rm sol}$,   the atmospheric $\Delta m^2_{\rm atm}$~\cite{Capozzi:2013csa} 
and sterile $\Delta m^2_{\rm st}$~\cite{Giunti:2013aea,Kopp:2013vaa} mass-squared differences. We have
$\Delta m^2_{\rm sol} \ll \Delta m^2_{\rm atm} \ll \Delta m^2_{\rm st} \sim {\mathcal O}(1)$~eV$^2$.
Here 
${\mathcal U}$
is the $4 \times 4$ active-sterile mixing matrix, parametrized as in~\cite{Mirizzi:2012we}, where
the values of the different mixing angles are 
given by the global fits of the active~\cite{Capozzi:2013csa} and of the sterile neutrino mixings~\cite{Giunti:2013aea,Kopp:2013vaa},
respectively. 
 The terms proportional to the Fermi constant $G_F$ in Eq.~(\ref{eq:omega})  are the standard matter effects in active neutrino oscillations.
In particular, the two contributions ${\sf E_\ell}$  and ${\sf E_\nu}$ are the energy density of $e^{\pm}$ pairs,  and $\nu$ and  $\bar\nu$, respectively.
 Finally, the term proportional to $G_X$  in Eq.~(\ref{eq:omega}) represents the new matter secret potential 
where ${\sf E_s}$ is  the energy density of $\nu_s$ and  $\bar\nu_s$. 

Flavor evolution generally occurs at  $T_\nu \ll M_X$ (we comment below for the case $T_\nu> M_X$), so that one can reduce it to a contact interaction, with an effective
strength~\footnote{See~\cite{Dasgupta:2013zpn} for an explicit calculation of the neutrino potential associated with
the secret interactions.}
\begin{equation}
G_X = \frac{\sqrt{2}}{8} \frac{g_X^2}{M_X^2} \,\ .
\label{eq:contact}
\end{equation}
The numerical factor $\sqrt{2}/8$ has been included in order to have in the $X$ sector the same relation which holds among 
the Fermi constant $G_F$, the $SU(2)_L$ coupling constant $g$ and the $W$-mass  in the Standard Model.
When this matter term is of the order of the vacuum oscillation frequency, associated with $\Delta m^2_{\rm st}$,
a MSW resonance  among active and sterile
neutrinos occurs~\cite{Matt}. 
In the following we will neglect the standard matter effects in neutrino oscillations, proportional
to $G_F$, since  we are in the limit $G_F \ll G_X$. 

The last term in the right-hand side of Eq.~(\ref{drhodt}) represents the collisional term.
Since we will work at $T_\nu \ll 1$~MeV, the standard collisional term $\propto G_F^2$ in the active
sector~\cite{Bell:1998ds} can be neglected and we only 
consider the collisional effect in the sterile sector, associated with the   secret self-interactions
$\nu_s \nu_s \to \nu_s \nu_s$, which reads~\cite{Chu:2006ua}
\begin{equation}
C[\rho] = {-i}\frac{\Gamma_X}{2}[{\sf S}_X, [\rho, {\sf S}_X]] \,\ ,
\label{eq:collis}
\end{equation}
 where
\begin{equation}
  \Gamma_X \simeq G_X^2 T_\nu^5 \frac{p}{\langle p \rangle} \frac{n_{s}}{n_a}\,\ ,
  \label{eq:dampgamma}
\end{equation}
 is the scattering rate~\cite{Bell:1998ds},%
~\footnote{Note a typo in the scattering rate   associated with secret interactions
 in~\cite{Dasgupta:2013zpn}, where it was written as $\propto  G_F^2$ instead of $G_X^2$. 
}
 with ${\langle p \rangle \simeq 3.15 \, T_\nu}$ the average-momentum for a thermal  Fermi-Dirac distribution, and
$n_s$ and $n_a$ the sterile and the active neutrino abundance, respectively.
In flavor basis, ${\sf S}_X = \textrm{diag}(0, 0,0,1)$ is the matrix
with the numerical coefficients for the scattering process.  Notice that, in the range we consider for $G_X$ the
elastic scattering terms which redistribute momenta are much larger than Hubble parameter, so we expect any initial sterile distribution will rapidly approach the standard Fermi-Dirac shape.

The interplay between flavor oscillations and collisions becomes more transparent considering the 
equations of motion 
[Eq.~(\ref{drhodt})--(\ref{eq:omega})]  in the case  of mixing of only one active 
flavor with the sterile neutrinos. In this case  one may write
$\rho= \frac{1}{2} (1 + {\bf P}\cdot \hbox{\boldmath$\sigma$})$,
${\sf\Omega} = \frac{1}{2} (\omega_0 + {\bf \Omega} \cdot \hbox{\boldmath$\sigma$})$
 and $ {\sf S}_X = \frac{1}{2} (s_0 + {\bf S}_X \cdot \hbox{\boldmath$\sigma$})$, where $\sigma$ are the Pauli matrices. 
One  recovers the known evolution of the polarization vector ${\bf P}$, expressed by the 
 Stodolsky's formula~\cite{Stodolsky:1986dx,Hannestad:2012ky}
\begin{equation}
\frac{d{\bf P}}{dt}= {\bf\Omega} \times {\bf P}- D\, {\bf P}_T \,\ .
\label{eq:eom2}
\end{equation}
 The first term at the right-hand-side represents the precession of the polarization vector {\bf P} around
${\bf \Omega}$.
The effect of the collisions is to destroy the coherence of
the flavor evolution, leading to a shrinking of the length of {\bf P}. More precisely, 
    the  component of the polarization vector  ``transverse'' to the flavor
basis ${\bf P}_T$ is damped with a rate
$D= (1/2) \Gamma_X |{\bf S}_X|^2$ (we remind the reader that ${\bf P}_T$  represents the off-diagonal elements of $\rho$). As shown in~\cite{Stodolsky:1986dx,Enqvist:1991qj}, the combination of precession  and damping can lead to different behaviors depending on the
relative strength
 of the two effects.

\begin{figure*}[!t]
 \includegraphics[angle=0,width=1.\columnwidth]{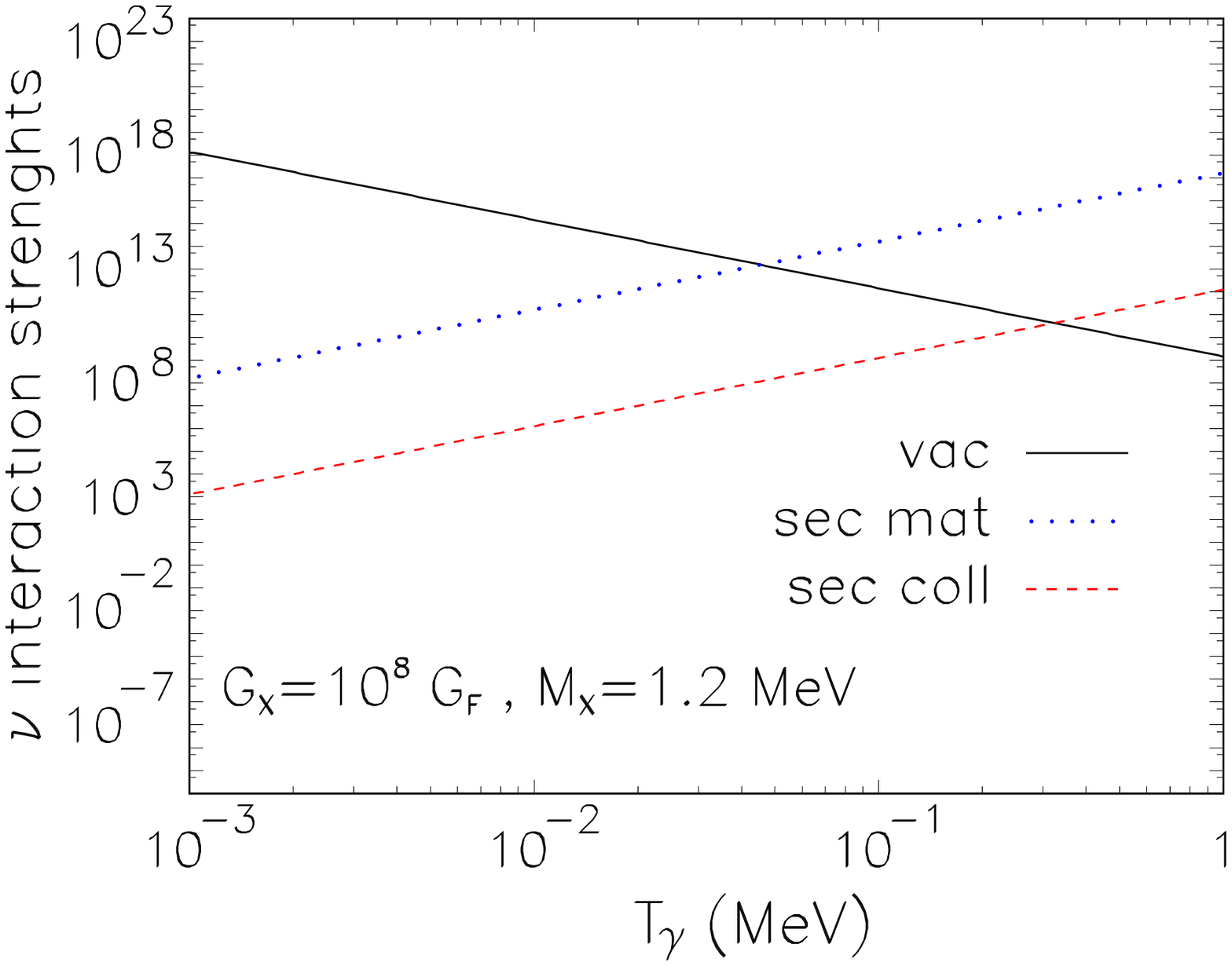} 
  \includegraphics[angle=0,width=1.\columnwidth]{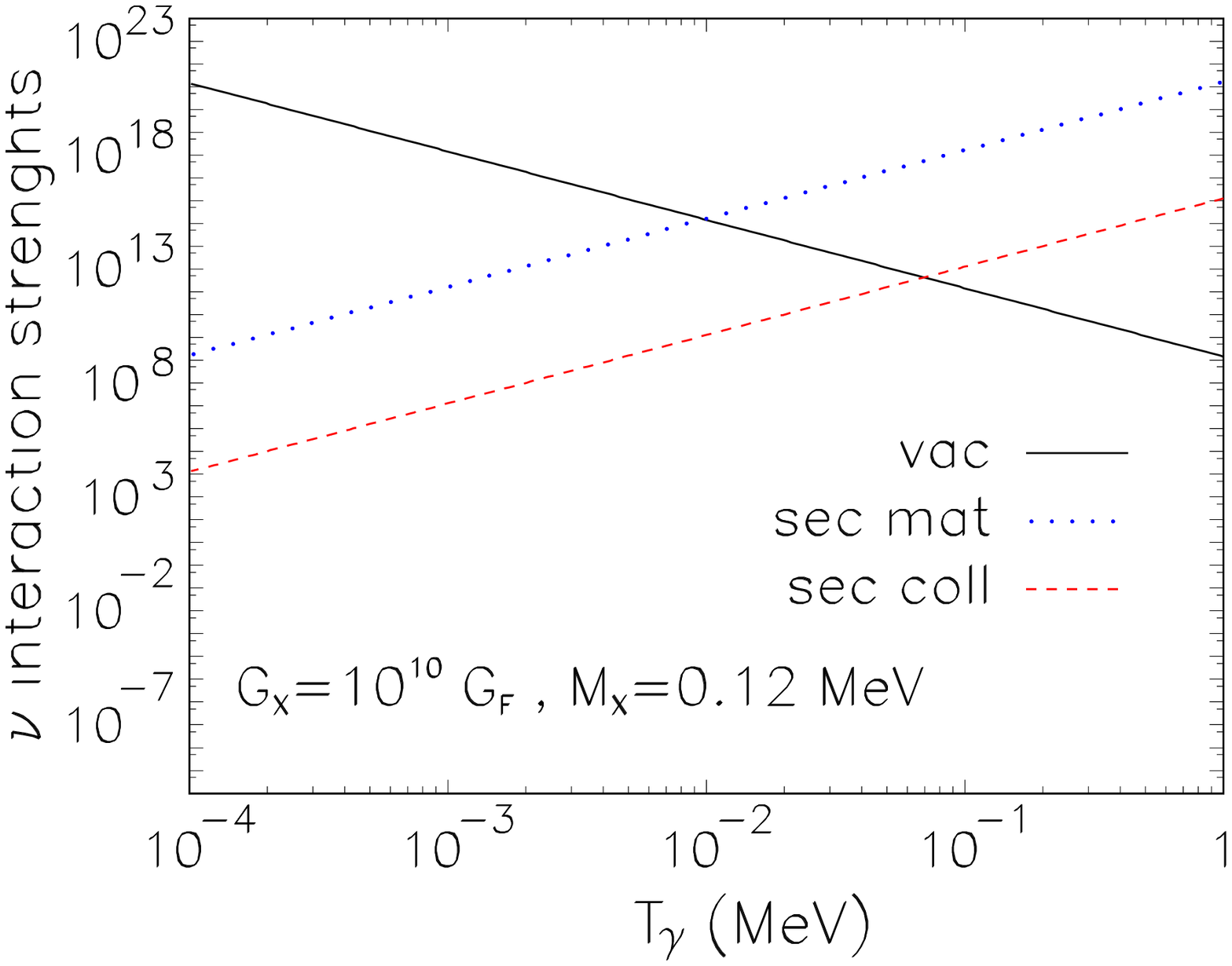} 
\caption{Neutrino  refractive and collisional rates (normalized in terms of the Hubble rate) versus photon temperature
 $T_{\gamma}$ for 
$g_X=0.1$. Left panel corresponds to $G_X=10^{8}$~$G_F$ and $M_X=1.2$~MeV, while right panel to
$G_X=10^{10}$~$G_F$ and $M_X=0.12$~MeV. 
The curves are the active-sterile vacuum  term (solid curve), the secret matter potential
for $\rho_{ss}=0.06$  (dotted curve), 
 and the scattering-induced decoherent production rate $\Gamma_t$ associated with $G_X^2$, assuming a sterile
neutrino abundance from vacuum oscillations (dashed curve)\label{fig1}.} 
\end{figure*}
   
In particular, when the typical
oscillation rate ($t_{\rm osc}^{-1} $),  collision rate ($t_{\rm coll}^{-1}$) and the expansion rate of the universe ($H$), obey the
hierarchy $t_{\rm osc}^{-1} \gg t_{\rm coll}^{-1} \gg H$
 the flavor dynamics can be described as follows. Active neutrinos $\nu_{\alpha}$ start to oscillate into sterile states $\nu_s$. 
The average probability to find a sterile neutrinos in a scattering time scale ($\Gamma_X^{-1}$) is given by
$\langle P(\nu_\alpha \to \nu_s) \rangle_{\rm coll}$. In each collision, the momentum of the $\nu_s$ component
is changed, while the $\nu_{\alpha}$ component remains unaffected. Therefore, after a collision the two
flavors are no longer in the same momentum state, and then  they can no longer oscillate.  Conversely,
they start to evolve
independently. However, the remaining active neutrinos can develop a new coherent  $\nu_s$
component which is made incoherent
in the next collision, and so on. The process continues till one reaches a flavor equilibrium with equal number
of $\nu_{\alpha}$ and $\nu_s$ (corresponding to $|{\bf P}|=0$, i.e. a completely mixed ensemble).    
Starting with a pure active flavor state, corresponding to 
$\rho=\textrm{diag}(1,0)$, the final density matrix
would be $\rho=\textrm{diag}(1/2,1/2)$.
The
averaged relaxation
rate to reach this chemical equilibrium is~\cite{Kainulainen:1990ds,Stodolsky:1986dx}
\begin{equation}
\Gamma_t \simeq \langle P(\nu_\alpha \to \nu_s) \rangle_{\rm coll}  \Gamma_X \,\ .
\label{eq:rel}
\end{equation}
 Notice that this rate is non zero as soon as an initial sterile neutrino density is produced when the matter term becomes of the order of the vacuum oscillation frequency. This initial sterile abundance is again proportional to the conversion probability. Thus  $\Gamma_t$ is,  at first, proportional to the square of $\langle P(\nu_\alpha \to \nu_s) \rangle_{\rm coll}$.

In the following we will show that the conditions to trigger this dynamics are always fulfilled at
$T_\nu \ll 1$~MeV, leading to a copious sterile neutrino production.
The key observation is that, although the 
refractive potential is smaller than the oscillation rate and the collisional cross sections are even smaller, yet they  exceed the Hubble rate and lead to  scattering-induced decoherent
production of sterile neutrinos.

 \section{Sterile neutrino production by scattering-induced decoherence}
In Fig.~\ref{fig1} we show the behavior of the different neutrino  refractive and collisional rates normalized to the Hubble rate $H(T_\gamma)$, versus photon temperature $T_\gamma = 1.4 T_\nu$
(see~\cite{Mirizzi:2012we} for details).
 For the sake of illustration, we show the quantity of Eq.~(\ref{eq:omega})-(\ref{eq:dampgamma}) averaged over thermal Fermi-Dirac distributions.
Results are shown for
$g_X=10^{-1}$. Left panel is for  $G_X=10^{8}$~$G_F$, or $M_X=1.2$~MeV, while right panel corresponds to
$G_X=10^{10}$~$G_F$, i.e. $M_X=0.12$~MeV. 
We show the active-sterile vacuum  term (solid curve) and the secret matter potential (dotted curve)
assuming $\rho_{ss}= n_s/n_a=0.06$, corresponding to the initial
sterile neutrino abundance induced by vacuum oscillations (see later).
We note that for $T_\nu>M_X$ the real form of the matter potential would deviate from the contact
 structure of  Eq.~(\ref{eq:contact})
used in the Figure (see~\cite{Dasgupta:2013zpn}). In particular, for $T_\nu \simeq M_X$ the potential would vanish,
leading to a possible production of $\nu_s$  when this condition is fulfilled.  However, 
since the duration of this phase is expected to be shorter than the 
inverse of the sterile neutrino production rate $\Gamma^{-1}_t$, 
for simplicity
we neglect this possible (small) extra-contribution of sterile neutrinos. 
In the left panel 
 a resonance would take place at $T_\gamma\simeq 5 \times 10^{-2}$~MeV, while in the right panel
at $T_\gamma\simeq 1 \times 10^{-2}$~MeV. 
This resonance excites sterile states. 

In principle one should perform numerical simulations in a (3+1) scheme in order to calculate the 
resonant sterile neutrino abundance and the further flavor evolution. However, in the presence of  the very large
matter potential and collisional term, induced by the secret interactions, these would be computationally demanding. 
Moreover, our main argument is not related to the details of the corresponding dynamics. Therefore, for simplicity we assume that the resonance 
 is completely non-adiabatic, so we have to take into account only the vacuum production of sterile neutrinos at lower temperatures when
 the matter term  becomes smaller than the vacuum oscillation term, associated with $\Delta m^2_{\rm st}$. This is a very conservative assumption. However, it allows us to easily compute the flavor evolution and is enough to show the role of the damping term.
 
The active-sterile vacuum oscillation probability, averaged over a collision time scale, is given by~\cite{Jacques:2013xr}
\begin{equation}
\langle P(\nu_\alpha \to \nu_s) \rangle_{\rm coll} \simeq \frac{1}{2}
\sin^2 2 \theta_{\alpha s}\,\ .
\end{equation}
Taking as representative mixing angle $\sin^2 2 \theta_{e s} \simeq  0.12$~\cite{Giunti:2013aea},
 one would expect a sterile neutrino abundance,  $n_{s} \simeq   0.06 \, n_a$. This seems a negligible contribution,
but is enough to generate a large  scattering rate proportional  to $G_X^2$, see Eq.~(\ref{eq:rel}). 
This is shown in Figure~\ref{fig1} as dashed curves.
As one can see, at $T_\gamma \lesssim 10^{-2}$~MeV,  $\Gamma_t\gg H(T_\gamma)$. 
Therefore, the  scattering-induced decoherent production will lead to a quick flavor equilibrium. 
Starting with a density matrix having
\begin{equation}
 (\rho_{ee}, \rho_{\mu \mu},\rho_{\tau \tau}, \rho_{ss})_{\rm initial}=(1,1,1,0)
\end{equation}
one thus, quickly reaches
\begin{equation}
(\rho_{ee}, \rho_{\mu \mu},\rho_{\tau \tau}, \rho_{ss})_{\rm final} =(3/4,3/4,3/4,3/4)
\end{equation} 
for all the parameter space
associated with eV sterile neutrino anomalies. 
Notice that this result
does not depend on the particular value of $G_X$, but only on the fact that the condition of strong
damping is realized.
The same equilibrium value would remain valid for example, for smaller masses $M_X$ by different order of magnitudes, at least until we can treat the collisional term as a four--point effective interactions, i.e. for $T_\nu> M_X$.
As discussed before, the rate of sterile neutrino re-thermalization is extremely fast, so that
the process is instantaneous.
Indeed, from Eq.~(\ref{eq:dampgamma}) and (\ref{eq:rel})  one can estimate $\Gamma_t \gtrsim 10^{-18}$~MeV for the cases
we have shown. This means that it is practically impossible to numerically follow the rise
of the sterile neutrino production. 
 However, it can be interesting 
to appreciate this dynamics in a case where the process is slower. 
We address the interested reader  to the lower panel of Fig.~3 in~\cite{Saviano:2014esa},
where it is shown the evolution of the diagonal elements of the density matrix $\rho$ in 
a (2+1) scheme, for a scenario with $g_X=10^{-2}$ and $G_X=10^{3} G_F$. In this case the sterile 
neutrino production  starts at $T\lesssim 1$~MeV and the final value would be $\rho=2/3$, as expected
with only three oscillating neutrino families.

Furthermore, we mention that the final equilibrium value does not depend on the exact values
of the active-sterile neutrino mixing angles. Indeed, we explicitly checked that also assuming that
the sterile species mixes only with an active one, all the flavors will participate to the 
equilibrium, due to the presence of the active mixing angles that connect the different species,
and indirectly link them also to the sterile species.

 Secret interactions mediated by a light (or even massless) pseudo-scalar ($M_X \ll T$), with a Lagrangian
${\mathcal L} \sim g_s X {\bar \nu} \gamma_5 \nu$, were considered in~\cite{Archidiacono:2014nda}.
It was found that couplings $g_s \sim 10^{-5}$ were sufficient to block thermalization
 prior to neutrino decoupling (and make dark matter sufficiently self-interacting).  However, as long as $g_s \gtrsim 10^{-6}$ the $\nu_s-X$ plasma
would be strongly interacting till sterile neutrinos become non-relativistic. Therefore, the 
mechanism of $\nu_s$ production and flavor equilibration would apply also to this case.
However, in the massless case, there would be also the production of a bath
of $X$ via the $\bar\nu_s \nu_s \to X X$ process (with a rate $\Gamma_X \sim g_s^4 T$). 
Therefore one would expect a chemical equilibrium over the five species, with an abundance $\rho=3/5$ for each of them.
As mentioned in~\cite{Archidiacono:2014nda}, a plasma of strongly-interacting $\nu_s$ and $X$ can have
interesting cosmological signatures, since the $\nu_s-X$ bath would behave as a single massless component
with no anisotropic stress.

 \section{Cosmological signatures}

\subsection{Effective neutrino species $N_{\rm eff}$}

The initial snapshot of active and sterile neutrino distribution soon after steriles are excited via oscillation is a shared grey body distribution, a Fermi-Dirac function weighted by a factor 3/4 for each species. In absence of any interaction, these distribution would remain frozen but for the effect of momentum redshift. In the model we are considering, after their production sterile neutrinos are fastly rescattering among themselves via secret interactions of order $G_X^2$. They are therefore, collisional. In fact, a grey body distribution is not a solution of the collisional Boltzmann equation, so these scattering processes will push the sterile distribution towards a Fermi-Dirac shape, with the constraint that the total neutrino number density is kept constant. Furthermore, as soon as sterile neutrinos  change their distribution this has a feedback on active neutrino distribution too, which are still efficiently oscillating into sterile states. This implies that all neutrino species in presence of sterile-sterile scattering will adjust quite efficiently their distribution to a thermal equilibrium distribution. The constant number density (or entropy) constraint implies that their eventual temperature  is reduced by a factor $ (3/4)^{1/3}$ with respect to the initial active neutrino temperature $T_\nu= (4/11)^{1/3} T_\gamma$. Indeed, we have
\begin{eqnarray}
&& n_\nu = 2 \int \frac{d^3 p}{(2 \pi) ^3} \frac34 \frac{1}{\exp(p/T_\nu)+1} \nonumber \\
&&= 2 \int \frac{d^3 p}{(2 \pi )^3}  \frac{1}{\exp[p/(T_\nu(3/4)^{1/3})]+1} \, .
\end{eqnarray}
There are in fact, no pair production processes from the electromagnetic plasma which can refill active neutrino densities, since in the scenario we are considering, sterile neutrinos are excited well below the active neutrino decoupling phase. 

From these considerations we see that the total energy density stored in active and sterile neutrinos is reduced. Indeed, before sterile neutrinos are excited,  and after $e^+-e^-$ annihilation phase, active neutrinos energy density is given by 
\begin{equation}
\epsilon_{\nu,\textrm{in}} = 3 \times  \frac{7}{8} \left( \frac{4}{11} \right)^{4/3} \epsilon_{\gamma} \,\ ,
\end{equation}
where the photon energy density is $\epsilon_{\gamma} = (\pi^2/15) T_\gamma^4$
and the effective number of neutrino species is $ N_{\rm eff} \simeq 3$, neglecting the small effect due to partial neutrino heating of order $\Delta N_{\rm eff}=0.046$ \cite{Mangano:2005cc}.
After sterile states are produced via oscillations and kinetic equilibrium is reached via secret interactions, we have four species which share a common temperature $T_\nu = (4/11)^{1/3} (3/4)^{1/3}T_\gamma$. 
Therefore, till all neutrinos are fully relativistic
\begin{equation}
\epsilon_{\nu,\textrm{fin}} =  4 \times \left( \frac{3}{4} \right)^{4/3}  \times\frac{7}{8} \left( \frac{4}{11} \right)^{4/3}  \epsilon_{\gamma}  \, ,
\end{equation}
and correspondingly the value of $N_{\rm eff}$ decreases to 
\begin{equation}
N_{\rm eff} \sim 4 \times \left(\frac34 \right)^{4/3} \, \sim 2.7 \, .
\end{equation}

This value is only slightly  reduced at the matter radiation equality, i.e. for $T_\gamma \sim 0.7$ eV, since at this energy scale only the high energy tail of sterile neutrino distribution counts as radiation. If we weight this contribution with the ratio of corresponding pressure over the pressure of a purely relativistic gas, as in \cite{Jacques:2013xr} and assume relativistic active states at this epoch, we find
with $m_{\rm st} \sim \sqrt{\Delta m^2_{\rm st}} \simeq 1$~eV
\begin{equation}
N_{\rm eff} \sim    3 \left(\frac34 \right)^{1/3}  \left( \frac34 + \frac14 \frac{P}{P_0} \right)  \sim 2.66  \, , \label{finalneff} \end{equation}
where the first and second terms in bracket are the active and sterile contribution, respectively and
\begin{eqnarray}
&& P = 2 \int \frac{d^3p}{(2 \pi)^3} \frac{p^2}{3 \sqrt{p^2+ {m_{\rm st}}^2}} \nonumber \\ 
&& \times \frac{1}{\exp[p/(T_\nu(3/4)^{1/3})]+1}  \, ,\\
&& P_0 = 2 \int \frac{d^3p}{(2 \pi)^3} \frac{p}{3} \frac{1}{\exp[p/(T_\nu(3/4)^{1/3})]+1}\, .
\end{eqnarray}
A value of the effective number of neutrino smaller than the expected standard result would be a signature in favour of strongly interacting sterile states mixed with active neutrinos, though there might be different models which can account for this result
(e.g., low-reheating scenarios~\cite{Ichikawa:2005vw}). Presently, the most precise determination of $N_{\rm eff}$ is from Planck experiment. In the standard $\Lambda$CDM model but allowing for a free number of relativistic species the result is  $N_{\rm eff} =3.30 \pm 0.27$ (68 \% C.L.) ~\cite{Ade:2013zuv}, which is compatible with (\ref{finalneff}) at about 2$\sigma$. A new data release by Planck collaboration is expected soon, including polarization data. This might put further constraint on $N_{\rm eff}$. In the future even more tight bounds are foreseen to come by next generation experiment such as Euclid~\cite{Laureijs:2011gra},which should reach a sensitivity of order $\Delta N_{\rm eff} < 0.1$.

 \subsection{Cosmological mass bounds}
 
The sterile neutrino production due to the  scattering-induced decoherent effects is expected to affect
 the Cosmic Microwave Background (CMB) and
Large Scale Structures (LSS), which are both sensitive to neutrino mass scale in the $10^{-1}$ eV-- 1 eV range.
One of the main effect of a massive neutrino is due their free streaming  till the epoch when they become non relativistic, which suppresses the growth of perturbations on small scales. However, if sterile states scatters via  secret interactions, the free streaming regime is delayed until the scattering rate becomes smaller than the Hubble parameter. This means that if $G_X$ is large enough so that this condition holds at the non relativistic transition, sterile neutrinos would never have a free streaming phase, but always diffuse.~\footnote{We are very pleased to thank Basudeb Dasgupta for pointing us this possibility.} The smaller value of $G_X$ for which this happens can be obtained from the condition that scattering rate equals the value of $H$ at a temperature $3.15 \,T_\nu \sim \langle p \rangle \sim \sqrt{\Delta m^2_{\rm st}}$ 
\begin{equation}
G_X^2 T_\nu^5 \sim H (T_\gamma) \, .
\end{equation}
Using standard expression of the Hubble rate in the $T_\gamma \sim$ eV range and the fact that, as we have seen, sterile temperature is given by  $T_\nu = (4/11)^{1/3} (3/4)^{1/3} T_\gamma$ this gives $G_X \sim  10^{10} G_F$, which corresponds to $M_X \simeq 10^{-1}$~MeV for  $g_X \simeq 10^{-1}$. Therefore, the mass bound discussed below only applies {\it as long as the coupling $G_X$ is smaller than this value}.

Assuming that the active neutrinos are much lighter than the sterile species, 
one can define an effective sterile neutrino mass~\cite{Mirizzi:2013kva}
\begin{equation}
m^{\rm eff}_{\rm st} =  \rho_{ss} \sqrt{\Delta m^2_{\rm st}} = \frac{3}{4} \sqrt{\Delta m^2_{\rm st}} \,\ .
\label{eq:effmass}
\end{equation}
 Latest analysis in~\cite{Giunti:2013aea}
of the sterile neutrino  anomalies
gives a best-fit $\Delta m^2_{\rm st} = 1.6$~eV$^2$ with a 
 $2 \sigma$ range 
\begin{equation}
1.08\, \ \textrm{eV}^2 < \Delta m^2_{\rm st} < 1.99 \,\ \textrm{eV}^2 \,\ .
\label{eq:giunti}
\end{equation}
Using Eq.~(\ref{eq:effmass})
 the lower value in 
the 2$\sigma$ range gives  $m^{\rm eff}_{\rm st} \simeq 0.78$~eV.
This value has to be compared with the cosmological mass bounds.
We also comment that the global analysis of sterile neutrino anomalies
presented in~\cite{Kopp:2013vaa} finds a discrepancy 
 between appearance and disappearance sterile neutrino  data. As a consequence
 only a small region around $\Delta m^2_{\rm st} \simeq 0.9$~eV$^2$ would be compatible
with all data.  This would correspond to  $m^{\rm eff}_{\rm st} \simeq 0.7$~eV.

Cosmological bounds on sterile neutrino mass and
abundance in the early universe are rather sensitive to
the data set used in the analysis, notably
CMB  data from Planck~\cite{Ade:2013zuv} and the recent but controversial BICEP2 experiment~\cite{Ade:2014xna} (see also~\cite{Adam:2014bub}), LSS, $H_0$ measurements as well as lensing and cluster data (CFHTLenS+PSZ).
In particular, the Planck Collaboration  combines  Planck  with WMAP polarization data, Baryon
Acoustic Oscillation and high multipole CMB data. In this case the  bound obtained is $m^{\rm eff}_{\rm st}< 0.42$~eV at 95$\%$ C.L.~\cite{Ade:2013zuv}, which is in strong disagreement with the sterile neutrino abundance produced by the secret interactions.
Moreover  a possible
non-zero sterile neutrino mass has been claimed in order to relieve the discrepancy between the CMB measurements and other observations, like current expansion rate $H_0$, the galaxy shear
power spectrum and counts of galaxy clusters, providing
 a value  $m^{\rm eff}_{\rm st} \simeq 0.7$~eV at $2\sigma$~\cite{Wyman:2013lza,Hamann:2013iba,Battye:2013xqa,Archidiacono:2014apa}
(see also~\cite{Battye:2014qga}) or  an upper bound of $m^{\rm eff}_{\rm st}
< 0.6$~eV~\cite{Leistedt:2014sia}. Stronger bounds have also be quoted in~\cite{Archidiacono:2014apa,Leistedt:2014sia}.

 In general, from the results  presented here, one would conclude that the minimum $m^{\rm eff}_{\rm st} \simeq 0.78$~eV
obtained
from the secret collisional production and compatible with the  $\Delta m^2_{\rm st}$ range
would be in tension (at least at 2$\sigma$ level) with 
the bounds on sterile neutrino mass from cosmology. A possible way out to this result is to consider extremely high couplings, $G_X \geq 10^{10} G_F$, since in this case sterile-sterile scatterings are in equilibrium till the eV scale, when they become non relativistic. As we mentioned, they would never experience a free streaming regime and cosmological mass bounds do not apply.

To close, we remark that in deriving our constraint we have been conservative, since
we have assumed that sterile neutrinos are produced only by vacuum oscillations 
at $T \ll 1$~MeV. However, it has been argued in~\cite{Dasgupta:2013zpn,Bringmann:2013vra} 
that there could be another colder primordial population of $\nu_s$
generated at $T \gg$~GeV by the decoupling of the $U(1)_X$ sector from standard particles. This additional contribution would increase the tension between the  sterile neutrino abundance  and  the cosmological mass bound.

\section{Conclusions}
Secret interactions among sterile neutrinos mediated by a light gauge boson $X$  have been recently 
proposed as an intriguing possibility to suppress the thermalization of eV sterile neutrinos in the early universe.
In particular, interactions mediated by a gauge boson with $M_X \leq 10$~MeV would suppress the sterile 
neutrino productions for $T\gtrsim 0.1$~eV and seemed therefore safe from cosmological constraints
related to big-bang nucleosynthesis~\cite{Saviano:2014esa}.

In the present work we have shown that when the matter potential produced by the 
sterile interactions becomes smaller than the vacuum oscillation frequency, sterile neutrinos are
copiously produced by the  scattering-induced decoherent  effects in the sterile neutrino sector. This process would lead  to a quick
flavor equilibration,  with a sterile neutrino abundance largely independent on the specific values of $G_X$ and $M_X$. 
A possible complete re-thermalization of sterile neutrinos and its impact on mass bound was already advocated 
in~\cite{Bringmann:2013vra}. Indeed, we have shown that due to the large damping effects this is \emph{always} the case in secret interactions among sterile neutrinos with low $M_X$ masses.

We investigated the cosmological consequences of this  huge sterile neutrino population.
We find that a signature of secret interactions would be a reduction of the effective number 
of neutrinos  $N_{\rm eff}$ down to 2.7.  If this value is compatible with the 2$\sigma$
range given by the  Planck experiment~\cite{Ade:2013zuv}, the future experiment Euclid~\cite{Laureijs:2011gra},
with a sensitivity to $\Delta N_{\rm eff} < 0.1$ may probe this small deviation with respect to the standard
expectation.
Moreover,
for $M_X \gtrsim g_X$~MeV,   sterile neutrinos would be free-streaming at the matter-radiation
equality epoch. Then, the  large sterile neutrino production would be in tension with the most recent cosmological mass bounds on sterile
neutrinos. 
We note that for the parameters $M_X \simeq g_X$~MeV, where
secret interactions would play also an interesting role in relation to dark matter and small-scale
structures~\cite{Dasgupta:2013zpn}, sterile neutrinos would be at the border between free-streaming
and collisional regime at the neutrino decoupling. In this case a dedicated investigations is necessary
to assess if mass bounds apply also in this case or can be evaded. 

Finally, we mention that recently it also been speculated that the background of sterile neutrinos produced by the collisions associated
with the secret interactions, would also modify the optical depth of ultra-high-energy recently observed by
 IceCube~\cite{Cherry:2014xra}.
Therefore,
future high-energy neutrino observations would be an interesting additional channel to probe this scenario.

\section*{Acknowledgements} 

 A.M. thanks Basudeb Dasgupta  
for useful comments on the manuscript and J{\"o}rn Kersten for useful discussions.
The work of A.M.   was supported by the German Science Foundation (DFG)
within the Collaborative Research Center 676 ``Particles, Strings and the
Early Universe.'' 
G. M., and O.P.  acknowledge support by
the {\it Istituto Nazionale di Fisica Nucleare} I.S. FA51 and the PRIN
2010 ``Fisica Astroparticellare: Neutrini ed Universo Primordiale'' of the
Italian {\it Ministero dell'Istruzione, Universit\`a e Ricerca}.
N.S. acknowledges support from
the European Union FP7 ITN INVISIBLES (Marie Curie
Actions, PITN- GA-2011- 289442).


\end{document}